\documentclass[prb,showpacs,amsmath,amssymb,twocolumn]{revtex4}

\usepackage{graphicx}
\usepackage[usenames]{color}
\usepackage[english]{babel}

\begin{document}

\title{High pressure Raman study of La$_{1-x}$Ca$_x$MnO$_{3-\delta}$ manganites}
\author{A. Sacchetti$^1$, T. Corridoni$^2$, E. Arcangeletti$^3$, P. Postorino$^3$}

\affiliation{$^1$Laboratorium f\"ur Festk\"orperphysik,
ETH-Z\"urich, CH-8093 Z\"urich, Switzerland\\
$^2$Dipartimento di
Fisica Universit\`a di ``Roma Tre'', Via della Vasca Navale 84,
I-00146 Roma, Italy\\
$^3$``Coherentia'' CNR-INFM and Dipartimento
di Fisica, Universit\`a di Roma ``La Sapienza'',  Piazzale Aldo
Moro 2, I-00185 Roma, Italy}

\begin{abstract}
We report on a high-pressure Raman study on two members of the
La$_{1-x}$Ca$_x$MnO$_{3-\delta}$ manganite family ($x=0.20$,
$\delta=0$ and $\delta=0.08$). The results obtained for the
$\delta=0$ sample show a different behavior in the low and high
pressure regimes ascribed to the onset of a new pressure-activated
interaction previously invoked in other manganite compounds. The
comparison of our results with literature data gives further
support to the identification of the Jahn-Teller active stretching
mode and shows that pressure-induced octahedral symmetrization is
more effective in systems exhibiting a lower metallic character.
On the contrary the new interaction sets in at pressure which
decreases on increasing the metallic character of the system
indicating a relevant role of the Mn-Mn hopping integral in its
activation.

\end{abstract}

\date{\today}

\maketitle

The peculiar properties of colossal magneto-resistive (CMR)
mixed-valence manganites\cite{MillisN,Dagotto} are commonly
described in the framework of double exchange
mechanism\cite{Zener} which competes with the localizing
electron-phonon coupling (EPC) triggered by the Jahn-Teller (JT)
distortion of the Mn$^{+3}$O$_6$ octahedra.\cite{Millis}
Nevertheless, the effects of the delicate balance among these
interactions on the macroscopic properties of these systems are
not yet completely understood.\cite{Dagotto} Moreover, in recent
years a number of experiments carried out on manganites under
pressure pointed out the relevance of a pressure-activated
localizing interaction which leads the system towards a new
unpredicted high pressure regime.
\cite{PRLold,Meneghini,PRLnew,FIR,Cinesi1,Cinesi2,Cinesi3,Cinesi4,Cinesi5,SacchettiCapone}

Usually, pressure-induced lattice compression in CMR manganites
remarkably affects the insulator to metal transition temperature,
$T_{IM}$, since the charge delocalization extent is directly
related to both the EPC strength and the hopping integral. In
principle, applying pressure results in an Mn-O-Mn bond length
compression and consequent linearization (i.e. an increase of the
hopping integral) and in a symmetrization the JT distorted MnO$_6$
octahedra (i.e. a reduction of the EPC). According to the above
prediction, early pressure-experiments showed an almost linear
increase of $T_{IM}$  within the $0-2$~GPa pressure
range,\cite{Radaelli2,Moritomo,Hwang2,Nuemeier,Kamenev,Wang,Lorenz}
whereas recent experiments carried out over much wider pressure
ranges showed that the observed low-pressure (LP) behavior cannot
be extended to the high-pressure (HP) regime. Indeed, pressure
becomes progressively less effective in increasing $T_{IM}$ and a
saturation regime where $T_{IM}$ is no more dependent on pressure,
is achieved \cite{PRLnew,Meneghini}. In several cases, when the
saturation regime sets in rather early (around 4-5~GPa), further
increase of pressure causes the opposite dependence with $T_{IM}$
starting to decrease
\cite{Cinesi1,Cinesi2,Cinesi3,Cinesi4,Cinesi5}.

Among the different investigated manganite compounds, a rather
complete set of high pressure (0$\sim$15~GPa) experimental data
(Raman,\cite{PRLold} Infrared,\cite{PRLnew} resistivity,
x-ray,\cite{Meneghini} and neutron\cite{Neutr} diffraction) is
available only for  La$_{0.75}$Ca$_{0.25}$MnO$_3$ (LC25S). In
particular, a HP Raman study of LC25S has shown a remarkable and
almost linear hardening of the peak frequency of the JT-active
stretching mode on increasing pressure up to 7~GPa, as expected
when the JT distortion is reduced. On the contrary, the peak
frequency remains almost constant on further increasing the
pressure up to 15~GPa.\cite{PRLold} A good agreement is found with
X-ray diffraction data which show a pressure-induced reduction of
the JT distortion over the LP regime. \cite{Meneghini} A
two-regime behavior for LC25S was also observed in temperature and
pressure dependent mid-infrared measurements aimed at determining
the insulator-to-metal transition curve. Indeed, $T_{IM}$
increases from 220~K  to $\sim$300~K going from zero to
$\sim$7~GPa but it keeps almost constant on further increasing the
pressure. Additional far-infrared measurements pointed out at the
failure of pressure in completely filling the insulating gap at
room temperature and in leading the system towards a coherent
transport regime.\cite{FIR} The whole of the data indicates the
onset of the new localizing mechanism which, at room temperature,
competes with the \textit{natural} charge-delocalizing tendency of
pressure and prevents both the full quenching of the JT distortion
and the metallization transition. Finally, the comparison between
the above experimental results and the theoretical calculations
presented in Ref.~\onlinecite{SacchettiCapone} suggests the
activation of an antiferromagnetic super-exchange coupling, which
is in conflict with the \textit{natural} pressure induced charge
delocalization, to be responsible for the anomalous high-pressure
behavior of LC25S.

In the present paper we focus on the effect of hole doping on the
high-pressure behavior of CMR La-Ca manganites to gain a deeper
understanding of the pressure effects and to find precursor
phenomena of the localizing mechanism. Hole doping, which converts
Mn$^{+3}$ into Mn$^{+4}$, can be varied by changing either
Ca-concentration $x$ or oxygen stoichiometry in
La$_{1-x}$Ca$_x$MnO$_{3-\delta}$ compounds. Indeed, since oxygen
is an electron acceptor, oxygen deficiencies reduce the number of
holes, which leads to an effective hole-doping
$x_{eff}=x-2\delta$. Owing to the different ionic radii of
La$^{3+}$ and Ca$^{2+}$ (see Ref.~\onlinecite{Shannon}), Ca-doping
induces a moderate reduction of the unit cell volume, whereas
oxygen-deficiency induces negligible structural
modifications.\cite{StrutDelta} Exploiting oxygen
non-stoichiometry it is possible to change the hole doping with
negligible structural effects, differently from what happens with
Ca-substitution.

We report on high-pressure Raman measurements on two samples of
La$_{0.80}$Ca$_{0.20}$MnO$_{3-\delta}$ with $\delta=0.00$ and
0.08, which corresponds to $x_{eff}=0.20$ and 0.04 respectively.
We remark that such a large hole-density variation is accompanied
by a small change in the unit cell volume (0.8\% according to
Ref.~\onlinecite{EminDelta}). The stoichiometric sample has the
same ground-state properties as LC25S, i.e. it is a ferromagnetic
metal below Curie temperature $T_C=194$~K, while the oxygen
reduced sample is an insulator at all temperatures, with a
ferromagnetic insulating phase below $T_C=163$~K. Preparation and
characterization of powder La$_{0.80}$Ca$_{0.20}$MnO$_{3}$ (LC20S)
and La$_{0.80}$Ca$_{0.20}$MnO$_{2.92}$ (LC20D) were described in
Ref.~\onlinecite{EminDelta}. High pressure far infrared
measurements on these sample were also reported in
Ref.~\onlinecite{FIR}.

Room temperature Raman spectra  were collected using a
confocal-microscope Raman spectrometer with the same experimental
setup and conditions as described in Ref.~\onlinecite{PRLold}. We
just recall that the low frequency cutoff of the notch filter
prevents the collection of reliable spectra below 200~cm$^{-1}$.
Samples was pressurized using a diamond anvil cell (DAC). The same
sample loading procedure as in Ref.~\onlinecite{PRLold} was
followed, where fine sample grains were placed on an NaCl pellet
pre-sintered in the DAC. This loading procedure ensures rather
good hydrostatic conditions and prevents laser-induced sample
heating \cite{PRLold} . At each pressure, four Raman spectra were
collected from different points of the sample, in order to average
over possible preferred orientations of the grains impinged by the
laser spot, $\sim$10~$\mu$m$^{2}$ on the sample surface in this
configuration.

\begin{figure}[tbp]
\includegraphics[width=6.5cm, angle=-90]{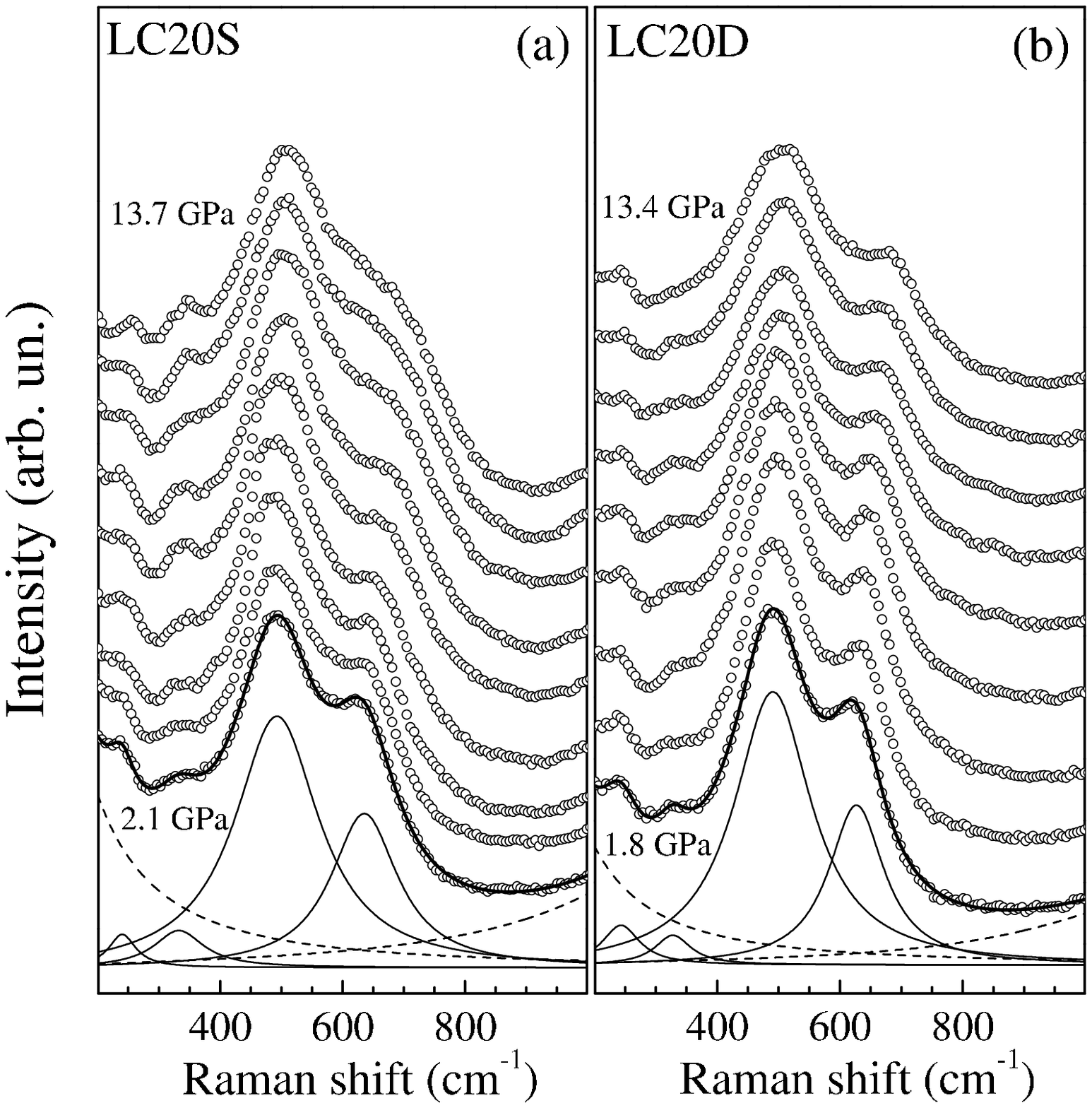}
\caption{Raman spectra of LC20S (a) and LC20D (b) at selected
pressures (open symbols). Data were progressively up-shifted
for clarity. Best-fit curves (thick
solid line) and fitting components (solid lines: phonons, dashed
lines: electronic and high frequency diamond contributions) are
also shown for both samples at the lowest pressure.}
\label{Spettri}
\end{figure}

\begin{figure}[!b]
\centering
\includegraphics[width=6cm, angle=-90]{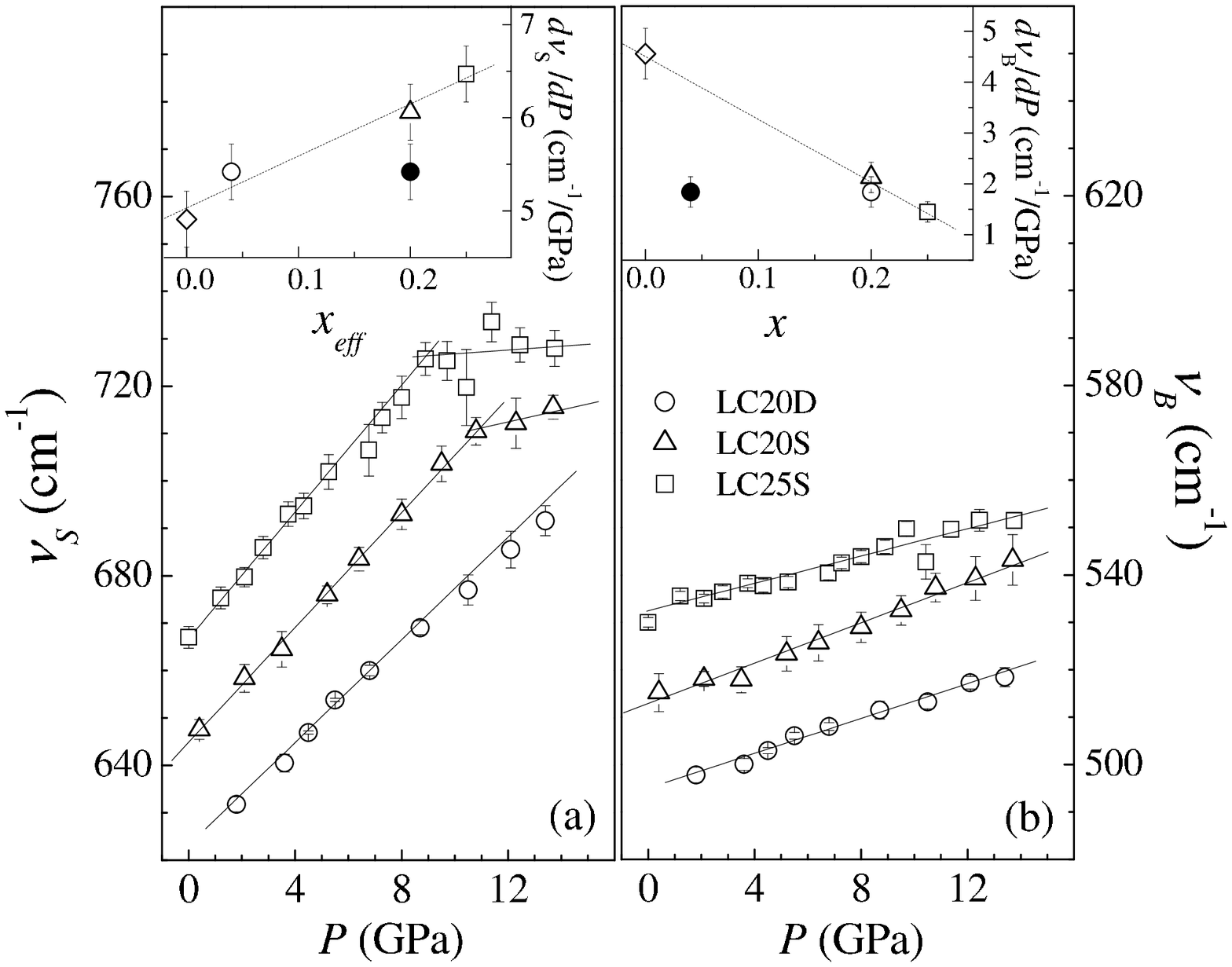}
\caption{Pressure dependence of phonon frequencies $\nu_S$ (a) and
$\nu_B$ (b) for LC20S, LC20D, and LC25S from
Ref.~\onlinecite{PRLold}. LC20D and LC25S data were up-shifted by
20~cm$^{-1}$ and 40~cm$^{-1}$ respectively. Insets: $d\nu_S/dP$
\textit{vs} effective doping, $x_{eff}$  (panel a); $d\nu_B/dP$
\textit{vs} Ca-doping, $x$ (panel b). Data for LaMnO$_3$
($x_{eff}=x=0$) are from Ref.~\onlinecite{Loa}. Pressure
derivatives for LC20D are also reported as a function of $x$ in
panel a) and $x_{eff}$ in panel b) (filled symbols). Solid/dashed
lines are guides to the eye.} \label{Fig2}
\end{figure}

Representative Raman spectra of LC20S and LC20D collected at
different pressure are shown in Fig.~\ref{Spettri} (a) and (b)
respectively.
All the spectra show four rather well defined phonon peaks: $\nu_1
\sim 250$~cm$^{-1}$, $\nu_2 \sim 330$~cm$^{-1}$, $\nu_3 \sim 490$
cm$^{-1}$, and $\nu_4 \sim 620$~cm$^{-1}$ at the lowest pressure.
These peaks can be assigned to the octahedron modes $A_g(2)$
($b$-axis rotation), $B_{3g}(4)$ ($c$-axis rotation), $A_g(3)$
(apical oxygen bending), and $B_{2g}(1)$ (in-plane oxygen
stretching), respectively.\cite{PRLold, Iliev} Although this
assignment is still debated,\cite{Iliev2}, in the following we
refer to the two peaks at the highest frequencies ($\nu_3$,
$\nu_4$ ) as bending ($\nu_B$) and stretching ($\nu_S$) phonons,
since further support to this assignment is here provided. Raman
spectra were fitted using the model curve\cite{PRLold}:
\begin{eqnarray}
S(\nu)= [1+n(\nu)] \bigg[\frac{A \nu \Gamma}{\nu^2+\Gamma^2}+
\nonumber
\\ \sum_{i=1}^5 \frac{A_i \nu
\Gamma_i}{(\nu^2-\nu_i^2)^2+\nu^2\Gamma^2}\bigg] \label{Fitfunc}
\end{eqnarray}
where $n(\nu)$ is the Bose thermal population factor, while the
first term in square brackets accounts for low-frequency diffusive
scattering from carriers with typical lifetime $\Gamma^{-1}$. The
linear combination of damped harmonic oscillators accounts for the
phonon contributions and for the broad structure at around
1100~cm$^{-1}$, due to the diamond fluorescence background. Good
fitting results were obtained for all the spectra using the model
of eq.~\ref{Fitfunc} (see Fig.~\ref{Spettri}). At each pressure,
the best-fit parameter values resulting from the analysis of the
spectra collected from the four zones were averaged and the
maximum dispersion value was taken as the data uncertainty. The
rather small dispersion found for the best-fit values of the
phonon frequency $\nu_i$ and linewidth $\Gamma_i$ shows that no
large pressure gradients are present. Owing to the low-frequency
cutoff, the relevant parameters for the electronic contribution
are affected by rather large uncertainties and do not show any
defined pressure dependence.  The frequency and the linewidth of
the two low-frequency phonons ($\nu_1$, $\nu_2$ and $\Gamma_1$,
$\Gamma_2$) remain constant within the uncertainty over the whole
pressure range, whereas the same quantities for bending and
stretching phonons ($\nu_B$, $\nu_S$ and $\Gamma_B$, $\Gamma_S$)
exhibit a remarkable pressure dependence. The pressure
dependencies of $\nu_S$ and $\nu_B$ are shown in Fig.~\ref{Fig2}
(a) and (b) for LC20S and LC20D, respectively, in comparison with
the corresponding data on LC25S from Ref.~\onlinecite{PRLold}.

A pressure induced hardening of $\nu_B$ and $\nu_S$ frequencies is
observed in all the three samples (see Fig.~\ref{Fig2}), with a
pressure rate much higher  for $\nu_S$ than for $\nu_B$. Moreover,
in LC20S and LC25S a two-regime behavior is well evident in the
$\nu_S$ pressure dependence (i.e. linear and almost pressure
independent at low and high pressure, respectively).

Within the LP regime, experimental $d\nu_{S}/dP$ and $d\nu_{B}/dP$
were obtained for LC20S, LC20D and compared with literature data
for LC25S \cite{PRLold} and for the parent compound LaMnO$_3$
\cite{Loa}. It is worth to notice that the pressure derivatives in
LaMnO$_3$ were obtained considering the Raman spectra up to about
8~GPa only, because the onset of a phase separation regime splits
the stretching peak into two components at higher
pressures\cite{Loa}. The pressure derivatives are shown in the
insets of Fig.~\ref{Fig2}: $d\nu_{S}/dP$ vs. effective doping
$x_{eff}$  and $d\nu_{B}/dP$ vs. Ca concentration $x$. We remark
that plotting the data vs $x$ or $x_{eff}$ shows a difference only
for the non-stoichiometric sample LC20D ($x\not=x_{eff}$). Undoped
LaMnO$_3$ ($x=x_{eff}=0$) shows nearly the same rate for $\nu_S$
and  $\nu_B$ whereas, on increasing the doping, $d\nu_S/dP$
increases and $d\nu_B/dP$ decreases. Moreover, focusing on the
comparison between LC20S ($x=x_{eff}=0.20$) and LC20D ($x=0.20$,
$x_{eff}=0.04$), it appears that $d\nu_B/dP$ depends on $x$ and
not on $x_{eff}$, since both the samples show the same rate
$d\nu_B/dP$, whereas $d\nu_S/dP$ exhibits a linear dependence only
when the data are plotted as a function of $x_{eff}$.

Bearing in mind that the number of Mn$^{+3}$ centered octahedra,
and thus the extent of JT distortion, is directly related to the
effective charge doping $x_{eff}$, the above findings support the
assignment of $\nu_S$ to a stretching mode strongly sensitive to
the JT distortion. Moreover, since the hardening of $\nu_S$
indicates local octahedra symmetrization,\cite{PRLold} the
$d\nu_S/dP$ behavior as a function of $x_{eff}$ indicates that, in
the LP regime, pressure is more effective in reducing the JT
distortion at high $x_{eff}$, due to the larger number of
undistorted Mn$^{+4}$ centered octahedra which make the lattice
somehow \textit{softer}. On the other hand, the pressure-induced
hardening of $\nu_B$ is consistent with the assignment of the
bending mode. On increasing $x$ the average ionic radius on the
rare-earth site is reduced, (La$^{3+}$ is larger than Ca$^{2+}$,
Ref.~\onlinecite{Shannon}) and the available space for bending the
apical oxygen ions increases. Therefore at large $x$, $\nu_B$ is
less affected by the pressure-induced lattice compression.

In the HP regime $d\nu_B/dP$ remains actually constant for all the
samples, whereas $d\nu_S/dP$ almost vanishes above 7~GPa and 9~GPa
for LC25S and LC20S respectively. No saturation effects were
observed for LC20D over the explored pressure range. The
difference between the LP and HP regimes is ascribed to the onset
of a new localizing mechanism competing with the pressure induced
symmetrization of the MnO$_6$ octahedron.\cite{Meneghini, PRLold}
Since the threshold pressure for the activation of the localizing
mechanism appears to decrease on increasing $x_{eff}$ , that is on
increasing the metallic character of the system, this finding
suggests that the new pressure activated interaction could be
somehow related to the effective Mn-Mn hopping integral, rather
than  to steric effects. These results are in agreement with far
infrared measurements\cite{FIR} on the same three samples showing
that, in samples with metallic ground state, the pressure-induced
charge-delocalization process is  much more pronounced in LC20S,
which exhibits a smaller $T_C=T_{IM}$ and a larger gap at $P=0$,
while at HP both LC20S and LC25S seem to approach the same regime.

\begin{figure}
\centering
\includegraphics[width=4.5cm, angle=-90]{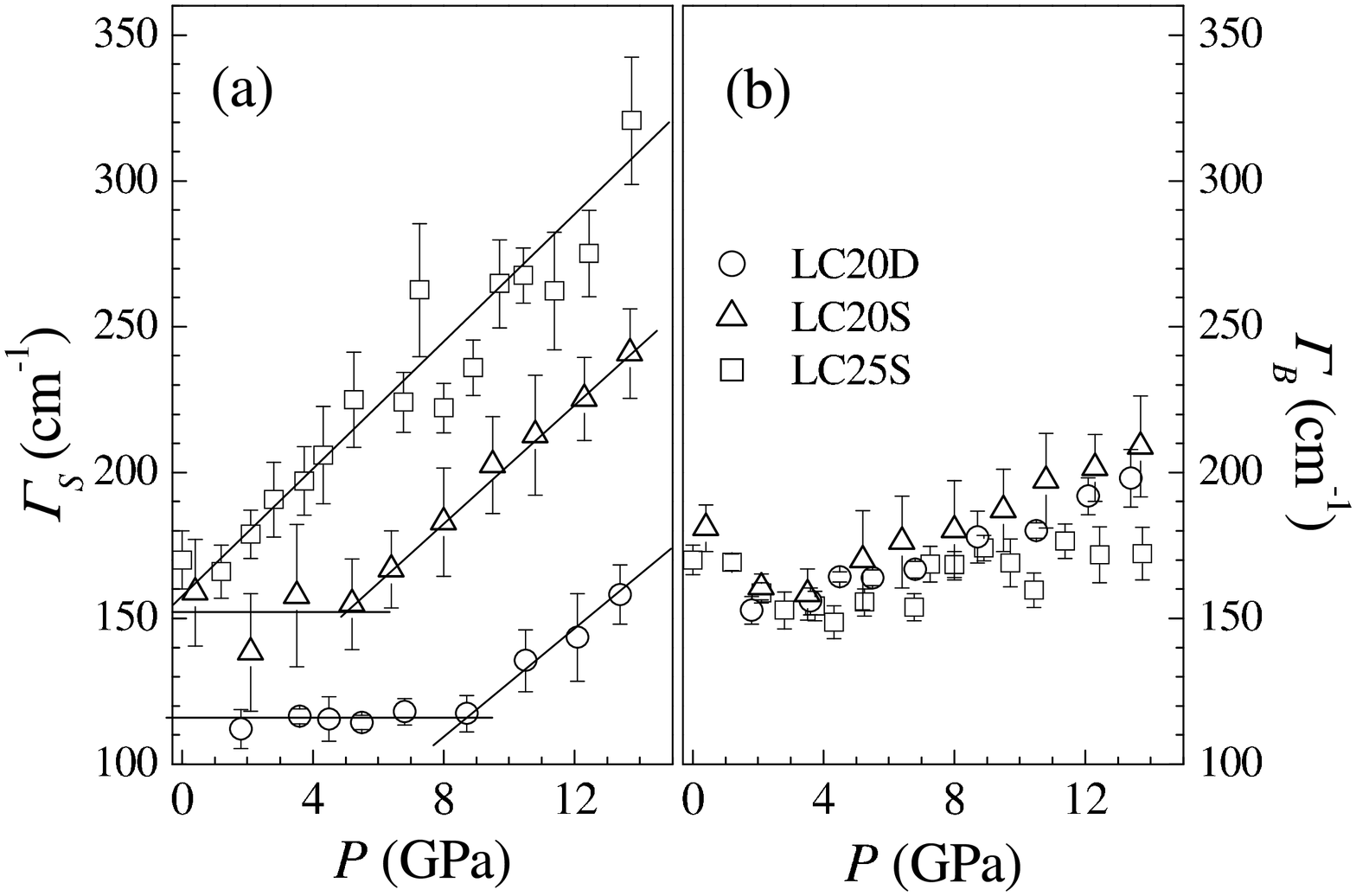}
\caption{Pressure dependence of the linewidth  $\Gamma_S$ (a) and
$\Gamma_B$ (b)  of the stretching and the bending modes
respectively, for LC20S and LC20D, compared with results on LC25S
from Ref.~\onlinecite{PRLold}. Solid lines are guides to the eye.}
\label{Fig3}
\end{figure}

The onset of the localizing mechanism can be seen also in the
pressure dependencies of the linewidths of the bending
($\Gamma_B$) and stretching ($\Gamma_S$) phonons of LC20S and
LC20D shown in Fig.~\ref{Fig3} (a) and (b), respectively, and
compared with corresponding results on LC25S from
Ref.~\onlinecite{PRLold}.
LC25S shows the largest $\Gamma_S$ and LC20D the smallest, whereas
$\Gamma_B$ is almost identical in all the three samples. This
effect can be interpreted in terms of the disorder of the JT
distortion induced by hole doping and it is more evident in LC25S
where $x$ and $x_{eff}$  are larger. The absence of any
appreciable difference in $\Gamma_B$ for the three doped samples
shows that the bending phonon is not correlated to carrier
density and thus to the JT distortion.

A  two-regime behavior can be envisaged for $\Gamma_S$: on
increasing pressure, at first the linewidth remains constant and,
above a threshold, it starts to increase linearly. A threshold
value of $\sim$ 8~GPa is rather apparent for LC20D whereas it can
be roughly evaluated at around 5~GPa for LC20S. In the case of
LC25S we can say that the threshold goes to zero and the linewidth
linearly increases starting from ambient pressure. The onset of
the lineshape broadening could be a precursor of the saturation
regime shown by the pressure dependence of $\nu_S$. On the other
hand, the linewidth behavior is also consistent with the onset of
a pressure-activated localizing interaction, which competes with
the \textit{natural} pressure-induced reduction of the JT
distortion. Indeed, the competition favors the coexistence of both
distorted and undistorted octahedra leading to a remarkable
disorder-induced broadening of the JT phonon. With this respect we
would like to recall that Raman measurements on LaMnO$_3$
($x=x_{eff}=0$, shows the onset of a regime characterized by the
appearance of domains of less JT distorted octahedra at $\sim$
8~GPa \cite{Loa}  that is at a pressure very close to the
threshold found for LC20D ($x_{eff}=0.04$).

In summary, we reported on high-pressure Raman measurements on
La$_{0.80}$Ca$_{0.20}$MnO$_{3-\delta}$ with $\delta=0.00$ and
0.08. A careful analysis of the phonon pressure dependence as a
function of Ca and effective charge doping provided further
support to the assignment of the bending and the JT-active
stretching octahedral modes. Moreover our results give
spectroscopic evidence of the onset of a localizing mechanism at
high pressure in metallic ground state manganites, confirming and
extending previous data on
La$_{0.75}$Ca$_{0.25}$MnO$_{3}$.\cite{PRLold} The comparison of
the present results with literature data allows us to evidence
that the greater the metallic degree of the system, the smaller
the pressure at which the localizing interaction effect sets in,
demonstrating that the strength of the new localizing interaction
depends on the effective Mn-Mn hopping integral, which increases
on increasing $x_{eff}$, as well as on increasing pressure. This
results is in agreement with the theoretical scenario proposed in
Ref.~\onlinecite{SacchettiCapone}, in which the localizing
super-exchange antiferromagnetic coupling becomes competitive with
double exchange at high pressure.

\end{document}